# Optimal designs for dose-finding experiments in toxicity studies

HOLGER DETTE[1], ANDREY PEPELYSHEV[2] and WENG KEE WONG[3]

[1]*Ruhr-Universität Bochum, Fakultät für Mathematik, 44780 Bochum, Germany.*
*E-mail: holger.dette@ruhr-uni-bochum.de*

[2]*St. Petersburg State University, Department of Mathematics, St. Petersburg, Russia.*
*E-mail: andrey@ap7236.spb.edu*

[3]*Departament of Biostatistics, University of California, Los Angeles, CA 90095-1772, USA.*
*E-mail: wkwong@ucla.edu*

We construct optimal designs for estimating fetal malformation rate, prenatal death rate and an overall toxicity index in a toxicology study under a broad range of model assumptions. We use Weibull distributions to model these rates and assume that the number of implants depend on the dose level. We study properties of the optimal designs when the intra-litter correlation coefficient depends on the dose levels in different ways. Locally optimal designs are found, along with robustified versions of the designs that are less sensitive to misspecification in the initial values of the model parameters. We also report efficiencies of commonly used designs in toxicological experiments and efficiencies of the proposed optimal designs when the true rates have non-Weibull distributions. Optimal design strategies for finding multiple-objective designs in toxicology studies are outlined as well.

## 1. Introduction

Developmental toxicity studies play an important role in identifying substances that may pose a danger to developing fetuses, including prenatal death and malformation among live fetuses. Krewski and Zhu (1995) and Zhu, Krewksi and Ross (1994) demonstrate that joint dose-response models for describing prenatal death and fetal malformation in developmental toxicity experiments have a good agreement with real data. These models can be used to estimate the effective dose corresponding to a gene excess risk for both these toxicological end-points, as well as for overall toxicity. It appears that toxicologists are generally less receptive to a more rigorous treatment of design issues; see a recent foreword/commentary in 2006 in *Nature* by Giles where the author expounded on the lack of sophistication in current designs for animal experiments. Very recently there have been a handful of theoretical articles that utilize statistical principles to design toxicology studies. This paper follows this trend and discusses how one may construct efficient designs for estimating malformation rate, prenatal death and overall toxicity levels under a broad range of model assumptions. A scientifically sound and efficient study is crucial because toxicology studies are increasingly more expensive in terms of time and labor. An efficient design potentially also means that significantly fewer animals will be required in the experiment. In what is to follow, our designs for such experiments are specified in terms of the number of doses to be used, the dose spacing, and the proportion of animals to be assigned to each dose.

Krewski, Smythe and Fung (2002) studied locally optimal designs for the estimating the effective dose using joint Weibull dose-response models. The locally optimal designs depend on the parameters of the Weibull model and the degree of intra-litter correlation. This paper addresses several important design issues in toxicology, such as estimating benchmark doses. Estimating benchmark doses has a long history in toxicological studies and research continues to this day; some recent papers include (Woutersen *et al.* (2001), Moerbeek *et al.* (2004), Slob *et al.* (2005)). As in Krewski, Smythe and Fung (2002), we seek optimal experimental designs that minimize the variance of the estimated effective dose for prenatal death, or malformation or overall toxicity, given the number of implants. (Here we recall an implant is an embryo that has been incorporated into the maternal uterus.) These are two important and common end-points measuring tetratogenicity (embryotoxicity) in animal studies (Zhu, Krewski and Ross (1994)). Because nonlinear models are employed in such studies, these authors used the simplest design strategy and found numerical locally optimal designs for estimating the model models. However, it is known that locally optimal designs can depend sensitively on the nominal or initial values of the model parameters. In our work, we present a more sophisticated approach to design developmental toxicity experiments so that the implemented design is more robust to misspecification in the initial values. Specifically, we first derive analytical properties of locally optimal designs for estimating the benchmark dose of prenatal death. In particular, we prove several results on the number and levels of doses and invariance properties of the locally optimal designs. We also correct an error in the work of Krewski, Smythe and Fung (2002), who used the wrong information matrix for the construction of the optimal designs. Second, we study the robustness properties of these locally optimal designs with respect to misspecification of the initial parameters. Third, we construct locally optimal designs for estimating the effective dose of overall toxicity and investigate their sensitivities to misspecification in the initial parameters. Fourth, we construct designs that are robust with respect to misspecification of the initial parameters, and so mitigate a concern raised by some toxicologists. We also investigate relative efficiencies of commonly used designs in developmental toxicity experiments.

Section 2 gives statistical background for our models, which were recently proposed in the literature for developmental toxicity studies. In Section 3 we present analytical results for locally optimal designs for estimating the effective dose of prenatal death and investigate the sensitivity of these designs with respect to misspecification of the unknown parameters. Section 4 considers similar problems for estimating the effective dose of overall toxicity, and in Section 5 the methodology is extended to obtain robust and efficient designs by a maximin approach. Section 6 evaluates efficiencies of commonly used designs in animal studies and briefly discusses efficiencies of optimal designs when non-Weibull probability models are used. All justifications for all our results are deferred to the Appendix.

## 2. Background for developmental toxicity studies

In developmental toxicity experiments with laboratory animals such as rats or mice, pregnant females are usually exposed to one of several doses of the test agent (including a control group at dose zero) during a specified period in gestation. Upon examining the uterine contents of each dam, the status of each conceptus is classified and recorded. A conceptus may either be dead or alive, and a live fetus may exhibit one or more malformations. In industry, the number of litters examined per dose is generally 24 for rats, which is one of the two standard species for segment II reproductive studies, at an estimated cost of $63,000 for the rat portion of the study. Rabbits are the other standard species in a segment II reproductive study and the number of rabbits required per dose is usually 12 with the cost of the study of this species being roughly $72,000 (Klaassen (2001), page 31–32).

Let $m_{ij}$ denote the number of implants in the $j$th litter of the $i$th dose $d_i$, and let $r_{ij}$ be the number of prenatal deaths, $s_{ij}$ be the number of live fetuses, and $y_{ij}$ be the number of fetal malformations. Summary observations from each dam yield a trinomial response $(r_{ij}, y_{ij}, s_{ij} - y_{ij})$ conditional on $m_{ij}$ for which we have

$$m_{ij} = r_{ij} + (s_{ij} - y_{ij}) + y_{ij}.$$

The fetal malformation rate $y_{ij}/s_{ij}$ and the prenatal death rate $r_{ij}/m_{ij}$ are of particular interest. The joint probability of the observed outcome $(y_{ij}, r_{ij}, m_{ij})$ may be factored as

$$\mathrm{P}(y_{ij}, r_{ij}, m_{ij}) = \mathrm{P}(y_{ij}|s_{ij}, m_{ij})\mathrm{P}(r_{ij}|m_{ij})\mathrm{P}(m_{ij}),$$

where $\mathrm{P}(m_{ij})$ is the marginal distribution of the implants number $m_{ij}$. Throughout, we let $\pi_1$ denote the probability of any malformation in a live fetus, $\pi_2$ be the probability of the prenatal death and let $\phi_i$ be the intra-litter correlation coefficient within $i$th dose group. Zhu, Krewski and Ross (1994) used generalized estimating equations in conjunction with an extended Dirichlet-multinomial covariance function, where the correlation coefficient is estimated separately. If $z_{ij} = (y_{ij}, r_{ij})^T$, the conditional covariance of the observation $z_{ij}$ is

$$\mathrm{Cov}(z_{ij}|m_{ij}) = m_{ij}(1 + (m_{ij} - 1)\phi_i)\begin{pmatrix} \mu(1-\mu) & -\mu\pi_2 \\ -\mu\pi_2 & \pi_2(1-\pi_2) \end{pmatrix},$$

where $\mu = \pi_1(1 - \pi_2)$, $1/(1 - m_{ij}) < \phi_i < 1$.

For simplicity we assume that $m_{ij}$ depends only on the dose level and not on the individual litter, that is, $m_{ij} = m_i = m(d_i)$. As pointed out in Krewski, Smythe and Fung

*Optimal designs for dose-finding experiments* 127(2002) this assumption avoids complicating the model with another level of estimation and permits the development of informative designs. Following Zhu, Krewski and Ross (1994) we use the Weibull model

$$\pi_i(d) = 1 - e^{-a_i - b_i d^{\gamma_i}}$$

to describe both probabilities $\pi_1$ and $\pi_2$, where $d$ denotes the dose level. Here $a_i$, $b_i > 0$ and $\gamma_i > 0$ are unknown parameters ($i = 1, 2$). We denote the parameters for the probability $\pi_i$ by $\theta_i$. Following Catalano *et al.* (1993) and Zhu, Krewski and Ross (1994), the *overall toxicity* is defined by

$$\pi_3(d) = 1 - (1 - \pi_1(d))(1 - \pi_2(d)) \tag{2.1}$$

of either a death or malformation occurring. The effective dose $ED_\alpha$ for a particular probability $\pi_i$ is defined as the (unique) solution of the equation

$$\frac{\pi(ED_\alpha) - \pi(0)}{1 - \pi(0)} = \alpha,$$

where $\pi(d)$ represents the probability of a response at dose $d$ and $\alpha$ is a given excess risk. The excess risk represents additional risk over background among animals that would not have responded under control conditions. Typical values for $\alpha$ are 0.01 or 0.05, and when $\alpha$ is set at a very low level, say $10^{-4}$, the excess risk is also sometimes called the benchmark dose or the virtually safe dose (Ryan (1992), Al-Saidy *et al.* (2003)). Zhu, Krewski and Ross (1994) proposed an estimate $\hat\theta$ for estimating $\theta$, the three parameters in the Weibull distribution. This estimate is based on quadratic estimating equations and has been shown to have reasonable efficiencies for estimating $\theta$. By the $\delta$-method (Van der Vaart (1998)) the variance of the estimator $\widehat{ED}_\alpha$ for the effective dose can be approximated by

$$\operatorname{Var}(\widehat{ED}_\alpha) \approx \tilde{D}^T \operatorname{Cov}(\hat\theta) \tilde{D}, \tag{2.2}$$

and

$$\tilde{D} = \frac{\partial}{\partial \theta} ED_\alpha \tag{2.3}$$

is the gradient of $ED_\alpha$ with respect to $\theta$. We will denote the vector of parameters in $\pi_i$ by $\theta_i$ and its corresponding estimate by $\hat\theta_i$, $i = 1, 2$.

Throughout, a design is specified by the number of different dose levels, say $k$, the specific dose levels $d_1, \ldots, d_k$ and the proportion of patients, say $w_1, \ldots, w_k$ allocated at each of these dose levels. In this paper, we consider approximate designs, that is, probability measures $\xi = \{d_i, w_i\}_{i=1}^k$ with finite support (Silvey (1980), Pukelsheim (1993)). For a given design $\xi$ and total sample size $n$, the number of observations at each dose level $n_j$ is obtained by rounding the quantities $nw_j$ to integers, such that $\sum_{j=1}^k n_j = n$ (Pukelsheim and Rieder (1992)). Throughout, we assume for the sake of simplicity that the dose range is given by the interval $[0, 1]$, but the adaption of the methodology to other dose intervals is straightforward. In what is to follow, we will only present our design strategy and results for estimating the effective dose of prenatal death. The strategy for estimating the effective dose for a given malformation rate is completely analogous and we omit details and corresponding results for space considerations.

## 3. Optimal designs for estimating the effective dose of prenatal death

Under the Weibull model the effective dose for prenatal death equals

$$ED_\alpha = \left(-\frac{\ln(1-\alpha)}{b_2}\right)^{1/\gamma_2}.$$

Recalling that $\theta_2^T = (a_2, b_2, \gamma_2)$, the gradient (2.3) in the representation (2.2) is given by

$$\tilde{D} = \frac{\partial}{\partial \theta_2} ED_\alpha = -\frac{ED_\alpha}{\gamma_2} \begin{pmatrix} 0 \\ 1/b_2 \\ \ln(-\ln(1-\alpha)/b_2)/\gamma_2 \end{pmatrix}$$

$$= -\frac{(-\ln(1-\alpha)/b_2)^{1/\gamma_2}}{\gamma_2} \begin{pmatrix} 0 \\ 1/b_2 \\ \ln(-\ln(1-\alpha)/b_2)/\gamma_2 \end{pmatrix}.$$

If $\xi = \{d_1, d_2, \ldots, d_n; w_1, w_2, \ldots, w_n\}$ denotes an approximate design and

$$D_i = \frac{\partial}{\partial \theta_2} \pi_2(d_i) = (1 - \pi_2(d_i)) \begin{pmatrix} 1 \\ d_i^{\gamma_2} \\ b_2 d_i^{\gamma_2} \ln(d_i) \end{pmatrix},$$

it follows that the covariance matrix of the estimate $\hat{\theta}_2$ is approximately

$$\mathrm{Cov}(\hat{\theta}_2) \approx M^{-1}(\xi, \theta_2),$$

where

$$M(\xi, \theta_2) = \sum_{i=1}^n w_i \frac{D_i D_i^T}{\mathrm{Var}(r_i \mid m_i)}$$

$$= \sum_{i=1}^n w_i \frac{D_i D_i^T}{m_i(1 + (m_i - 1)\phi_i)\pi_2(d_i)(1 - \pi_2(d_i))} \quad (3.1)$$

is the information matrix of the design. Note that the summands in this matrix differ by the factors $m_i^2$ from the corresponding terms in the information matrix derived by Krewski, Smythe and Fung (2002). There is, in fact, an error in that paper. Consequently we obtain from (2.2) as first-order approximation for the variance of the estimate of the effective dose

$$\mathrm{Var}(\widehat{ED}_\alpha) \approx \Psi(\xi, \theta_2) = \tilde{D}^T M^{-1}(\xi, \theta_2)\tilde{D} \tag{3.2}$$

and a locally optimal design for estimating the effective dose (of prenatal death) minimizes the function $\Psi$ among all designs for which the $ED_\alpha$ is estimable.

It is clear that the information matrix of the optimal design depends on the parameters of the model, and, in particular on the quantities $m_i = m(d_i)$ and $\phi_i = \phi(d_i)$. The simplest way to deal with this added complication is to use locally optimal designs proposed by Chernoff (1953). This strategy requires that a single prior guess for the unknown parameters is available. In developmental toxicity experiments such knowledge is often available from preliminary studies. The following results establish properties of locally optimal designs for estimating the effective dose of prenatal death. In essence, it says that if the excess risk is not too extreme (i.e., near 0 or 1), the locally optimal design requires only two doses; otherwise the locally optimal design requires three doses that include the extreme levels in the dose interval. The proofs rely on the geometric characterization of $c$-optimal designs of Elfving (1952) and are deferred to the Appendix.

**Theorem 1.** *Let $m$ and $\phi$ denote the functions defining $m_i = m(d_i)$ and $\phi_i = \phi(d_i)$. If the function*

$$d \longrightarrow \frac{(1 - \pi_2(d))}{m(d)(1 + (m(d) - 1)\phi(d))\pi_2(d)} \tag{3.3}$$

*is decreasing, there exist numbers $\underline{\alpha}$ and $\bar{\alpha}$ such that the following properties hold:*

(a) *If $\alpha \in (0, \underline{\alpha}] \cup (\bar{\alpha}, 1)$, the locally optimal design for estimating the effective dose of prenatal death is supported at three points, including the boundary points $d_1^* = 0$ and $d_3^* = 1$ of the design space.*

(b) *If $\alpha \in (\underline{\alpha}, \bar{\alpha})$, the locally optimal design for estimating the effective dose of prenatal death is supported at two points.*

We note that if the function $\phi$ is increasing, the assumption of Theorem 1 is satisfied. In particular, Bowman, Chen and George (1995) proposed a logistic-type function of the form

$$\phi(d) = \frac{2}{1 + \mathrm{e}^{u_1 + u_2 d}} - 1 \tag{3.4}$$

for describing the relationship between intra-litter correlation and dose, which is widely used in practice. If $u_2 < 0$ this function satisfies the assumptions of Theorem 1. In what is to follow, we will use this function to model the intra-class correlation using different values of $u_1$ and $u_2$. Of course the case of zero or no correlation is assumed when we set $u_1 = u_2 = 0$.

The next result tells us that we can limit our search for the optimal design to the protocol interval $[0, 1]$ and deduce the corresponding optimal design on the design interval $[0, T]$.

**Theorem 2.** *Assume that the quantity $m_i$ and the function $\phi$ are constant. The weights of the locally optimal design for estimating the effective dose of prenatal death do not depend on the parameter $\gamma_2$. Moreover, if $d_i^*(a_2, b_2, \gamma_2, T)$ are the support points of the locally optimal design for estimating the effective dose of prenatal death on the interval $[0, T]$, we have*

$$d_i^*(a_2, b_2, \gamma_2, 1) = (d_i^*(a_2, b_2, 1, 1))^{1/\gamma_2},$$
$$d_i^*(a_2, b_2, \gamma_2, T) = T d_i^*(a_2, b_2 T^{\gamma_2}, \gamma_2, 1).$$

Our next result shows that if the locally optimal design for estimating the prenatal death requires three dose levels, these dose levels do not depend on the value of the excess risk $\alpha$. It also provides a complete analytical description of the locally optimal design when it is known in advance that the locally optimal design needs only two doses, and one of them is the zero dose.

**Theorem 3.** *Assume that the conditions of Theorem 1 are satisfied.*

(a) *The support points of the locally optimal design for estimating the effective dose of prenatal death with three support points do not depend on the value of $\alpha$.*

(b) *If the support of a two-point locally optimal design for estimating the effective dose of prenatal death contains the point $0$, then the second support point is equal to $ED_\alpha$ and its weight at $ED_\alpha$ is equal to $w_2 = g(0)/(g(0) + g(ED_\alpha))$, where*

$$g(d) = \sqrt{\frac{(1 - \pi_2(d))}{m(d)(1 + (m(d) - 1)\phi(d))\pi_2(d)}}.$$

In Table 1 we display numerical locally optimal designs for estimating the effective dose of prenatal death for various combinations of the parameters when the quantity $m_i$ and the function $\phi$ are assumed to be constants. As stated in Theorem 1 the locally optimal designs are either two-point designs or three-point designs because the monotonicity assumption of the theorem is satisfied. In Table 2 we show some results for a non-constant function $\phi$ of the form (3.4), which demonstrate that the assumption of monotonicity on the function (3.3) is, in fact, needed. For example, if $\phi(d) = 2/(1 + e^{d-1}) - 1$ the corresponding function in (3.3) is not decreasing. The locally optimal design for estimating the effective dose of prenatal death is a 3-point design, but its support dose not contain the minimal dose 0.

In Table 1 and other tables that follow, we also display on the extreme right column the efficiency of an equally weighted design supported at five equally spaced points on the interval $[0, 1]$. We denote this design by $\xi_u$ and note that this is an example of a uniform design where doses are equally spread out and an equal number of observations are planned at each dose. Uniform designs are intuitively appealing because of their simplicity. In practice, approximately uniform designs are used. For example, a recent paper by Lee *et al.* (2006) used $0, 0.5, 1, 2, 3, 3.5$ (mg/kg) dose levels of cadmium, and separately, $1, 3, 5, 7.5$ dose levels of all-trans-retinoic acid to study their individual and combined effects on the induction of forelimb ectrodactyly in C57BL/6 mice. In the cases considered in both tables, the results show that this particular uniform design did not perform well, averaging about 50%. This means that roughly twice as many rats will be needed in the uniform design to obtain estimates for the parameters as accurate as those provided by the locally optimal design.

**Table 1.** Locally optimal design for estimating the effective dose of prenatal death conditional on the number of implants, assuming the functions $\phi$ and $m$ are constants. The table also shows the efficiency of the equidistant design $\xi_u = \{0, 1/4, 1/2, 3/4, 1; 1/5, 1/5, 1/5, 1/5, 1/5\}$ (last column)

| $\alpha$ | $a_2$ | $b_2$ | $\gamma_2$ | $d_1$ | $d_2$ | $d_3$ | $w_1$ | $w_2$ | $w_3$ | ED | eff($\xi_u$) |
|---|---|---|---|---|---|---|---|---|---|---|---|
| 0.05 | 0.13 | 0.15 | 3.33 | 0 | 0.725 |   | 0.455 | 0.545 |   | 0.725 | 0.506 |
| 0.05 | 0.13 | 0.2  | 3.33 | 0 | 0.696 | 1 | 0.429 | 0.545 | 0.025 | 0.665 | 0.539 |
| 0.05 | 0.13 | 0.25 | 3.33 | 0 | 0.689 | 1 | 0.404 | 0.547 | 0.049 | 0.621 | 0.557 |
| 0.05 | 0.13 | 0.3  | 3.33 | 0 | 0.682 | 1 | 0.387 | 0.549 | 0.064 | 0.588 | 0.568 |
| 0.05 | 0.13 | 0.35 | 3.33 | 0 | 0.676 | 1 | 0.373 | 0.551 | 0.075 | 0.562 | 0.575 |
| 0.05 | 0.13 | 0.4  | 3.33 | 0 | 0.670 | 1 | 0.363 | 0.554 | 0.083 | 0.540 | 0.579 |
| 0.05 | 0.01 | 0.27 | 3.33 | 0 | 0.607 |   | 0.285 | 0.715 |   | 0.607 | 0.481 |
| 0.05 | 0.05 | 0.27 | 3.33 | 0 | 0.653 | 1 | 0.371 | 0.593 | 0.036 | 0.607 | 0.541 |
| 0.05 | 0.1  | 0.27 | 3.33 | 0 | 0.678 | 1 | 0.390 | 0.558 | 0.051 | 0.607 | 0.558 |
| 0.05 | 0.15 | 0.27 | 3.33 | 0 | 0.690 | 1 | 0.399 | 0.543 | 0.058 | 0.607 | 0.564 |
| 0.05 | 0.2  | 0.27 | 3.33 | 0 | 0.698 | 1 | 0.404 | 0.535 | 0.061 | 0.607 | 0.568 |
| 0.05 | 0.25 | 0.27 | 3.33 | 0 | 0.703 | 1 | 0.407 | 0.529 | 0.063 | 0.607 | 0.570 |
| 0.03 | 0.13 | 0.27 | 3.33 | 0 | 0.686 | 1 | 0.367 | 0.538 | 0.095 | 0.519 | 0.590 |
| 0.04 | 0.13 | 0.27 | 3.33 | 0 | 0.686 | 1 | 0.381 | 0.543 | 0.076 | 0.567 | 0.577 |
| 0.05 | 0.13 | 0.27 | 3.33 | 0 | 0.686 | 1 | 0.396 | 0.548 | 0.056 | 0.607 | 0.562 |
| 0.06 | 0.13 | 0.27 | 3.33 | 0 | 0.686 | 1 | 0.412 | 0.554 | 0.034 | 0.642 | 0.546 |
| 0.07 | 0.13 | 0.27 | 3.33 | 0 | 0.686 | 1 | 0.430 | 0.560 | 0.010 | 0.674 | 0.526 |
| 0.08 | 0.13 | 0.27 | 3.33 | 0 | 0.703 |   | 0.433 | 0.567 |   | 0.703 | 0.507 |
| 0.1  | 0.13 | 0.27 | 3.33 | 0 | 0.754 |   | 0.420 | 0.580 |   | 0.754 | 0.499 |

**Table 2.** Locally optimal design for prenatal death conditional on the number of implants when the function $m$ is assumed to be constant and the function $\phi$ modeling the correlation is given by (3.4) ($a_2 = 0.13$, $b_2 = 0.27$, $\gamma_2 = 3.33$, $\alpha = 0.05$). The table also shows the efficiency of the equidistant design $\xi_u = \{0, 1/4, 1/2, 3/4, 1; 1/5, 1/5, 1/5, 1/5, 1/5\}$ (last column)

| $u_1$ | $u_2$ | $d_1$ | $d_2$ | $d_3$ | $w_1$ | $w_2$ | $w_3$ | ED | eff($\xi_u$) |
|---|---|---|---|---|---|---|---|---|---|
| 0  | −1 | 0     | 0.636 | 1 | 0.284 | 0.691 | 0.025 | 0.607 | 0.490 |
| 0  | −2 | 0     | 0.630 | 1 | 0.242 | 0.737 | 0.021 | 0.607 | 0.472 |
| 0  | −3 | 0     | 0.633 | 1 | 0.220 | 0.757 | 0.023 | 0.607 | 0.464 |
| −1 | 1  | 0.082 | 0.746 | 1 | 0.447 | 0.482 | 0.071 | 0.607 | 0.588 |
| −2 | 2  | 0.071 | 0.762 | 1 | 0.445 | 0.487 | 0.068 | 0.607 | 0.569 |
| −3 | 3  | 0.052 | 0.767 | 1 | 0.432 | 0.503 | 0.065 | 0.607 | 0.551 |

**Table 3.** Efficiency for estimating the effective dose of prenatal death. $\xi^*(\theta_0)$: locally optimal design for $\theta_0^T = (a_2, b_2, \gamma_2) = (0.13, 0.27, 3.33)$ $\xi_u$ equidistant design with five different dose levels (including the largest and smallest dose), and design $\xi_{mm} = \{0, 0.694, 1; 0.349, 0.515, 0.136\}$ which is standardized maximin optimal for estimating the effective dose of prenatal death with respect to $\Omega = [0.05, 0.2] \times [0.2, 0.4] \times [2.5, 4.5]$

| $a_2$ | 0.05 | 0.05 | 0.05 | 0.05 | 0.2 | 0.2 | 0.2 | 0.2 |
|---|---|---|---|---|---|---|---|---|
| $b_2$ | 0.2 | 0.2 | 0.4 | 0.4 | 0.2 | 0.2 | 0.4 | 0.4 |
| $\gamma_2$ | 2.5 | 4.5 | 2.5 | 4.5 | 2.5 | 4.5 | 2.5 | 4.5 |
| eff($\xi^*(\theta_0)$) | 0.802 | 0.766 | 0.526 | 0.967 | 0.944 | 0.695 | 0.749 | 0.862 |
| eff($\xi_u$) | 0.522 | 0.475 | 0.563 | 0.521 | 0.550 | 0.496 | 0.588 | 0.544 |
| eff($\xi_{mm}$) | 0.808 | 0.740 | 0.663 | 0.923 | 0.920 | 0.665 | 0.872 | 0.822 |

In general, our numerical results show that there are four types of locally optimal designs for estimating the effective dose of prenatal death, namely:

$$\{0, d_2, 1; w_1, w_2, w_3\}, \qquad \{0, d_2; w_1, w_2\}, \qquad \{d_1, d_2, 1; w_1, w_2, w_3\}, \qquad \{d_1, d_2; w_1, w_2\}.$$

Moreover, if the assumptions of Theorem 1 are satisfied, there exist only two types, that is,

$$\{0, d_2, 1; w_1, w_2, w_3\}, \qquad \{d_1, d_2; w_1, w_2\}.$$

Before any design is implemented, it is useful to investigate the robustness of the locally optimal designs for estimating the effective dose with respect to misspecification in the initial parameters. For this purpose we consider the locally optimal $\xi^*(\theta_0) = \{0, 0.686, 1; 0.396, 0.548, 0.056\}$ for the parameter $\theta_0^T = (a_2, b_2, \gamma_2) = (0.13, 0.27, 3.33)$ and calculate the efficiency

$$\text{eff}(\xi) = \frac{\tilde{D}^T M^{-1}(\xi, \theta) \tilde{D}}{\tilde{D}^T M^{-1}(\xi, \theta_0) \tilde{D}} \tag{3.5}$$

for various values of $\theta$. These results are listed in Table 3. We observe that locally optimal designs are not too sensitive with respect to changes of the parameter $a_2$, but a misspecification of the parameters $b_2$ and $\gamma_2$ has more serious effects. The table also shows the corresponding efficiencies of the equidistant design $\xi_u = \{0, 1/4, 1/2, 3/4, 1; 1/5, 1/5, 1/5, 1/5, 1/5\}$. In most cases these are smaller than the efficiencies of the locally optimal design for estimating the effective dose. In addition, the table contains efficiencies of a maximin design $\xi_{mm}$, whose construction will be motivated in Section 5. This design performs substantially better than the uniform design $\xi_u$ and achieves nearly the same efficiencies as the locally optimal design $\xi^*(\theta_0)$ in those cases where $\xi^*(\theta_0)$ is very efficient.

## 4. Dose-finding for overall toxicity conditional number of implants

If two Weibull models with parameters $\theta_1^T = (a_1, b_1, \gamma_1)$ and $\theta_2^T = (a_2, b_2, \gamma_2)$ are used for modeling the overall toxicity in (2.1), the effective dose based on $\pi_3(d)$ is defined as a solution of the equation

$$\alpha = 1 - \exp(b_1 ED_\alpha^{\gamma_1} + b_2 ED_\alpha^{\gamma_2}),$$

or, equivalently,

$$-\ln(1-\alpha) = b_1 ED_\alpha^{\gamma_1} + b_2 ED_\alpha^{\gamma_2}.$$

The approximation for the variance of the estimator based on generalized estimating equations is given by (2.2), where $\theta^T = (\theta_1, \theta_2)$ and

$$\bar{D} = \frac{\partial}{\partial \theta} ED_\alpha = \frac{-1}{b_1 \gamma_1 ED_\alpha^{\gamma_1 - 1} + b_2 \gamma_2 ED_\alpha^{\gamma_2 - 1}} \begin{pmatrix} 0 \\ ED_\alpha^{\gamma_1} \\ b_1 ED_\alpha^{\gamma_1} \ln(ED_\alpha) \\ 0 \\ ED_\alpha^{\gamma_2} \\ b_2 ED_\alpha^{\gamma_2} \ln(ED_\alpha) \end{pmatrix}.$$

If $\xi = \{d_1, d_2, \ldots, d_n; w_1, w_2, \ldots, w_n\}$ denotes an approximate design we have

$$\mathrm{Cov}(\hat{\theta}) \approx M^{-1}(\xi, \theta),$$

where the information matrix is given by

$$M(\xi, \theta) = \begin{pmatrix} M_1(\xi, \theta) & 0 \\ 0 & M_2(\xi, \theta) \end{pmatrix}$$

and the two non-vanishing blocks are defined by

$$M_1(\xi, \theta) = \sum_{i=1}^n w_i \frac{D_{(1)i} D_{(1)i}^T}{\mathrm{Var}(y_i \mid m_i)}$$

$$= \sum_{i=1}^n w_i \frac{D_{(1)i} D_{(1)i}^T}{m_i(1 + (m_i - 1)\phi_i)\pi_1(d_i)(1 - \pi_2(d_i))(1 - \pi_1(d_i)(1 - \pi_2(d_i)))},$$

$$M_2(\xi, \theta) = \sum_{i=1}^n w_i \frac{D_{(2)i} D_{(2)i}^T}{\mathrm{Var}(r_i \mid m_i)}$$

$$= \sum_{i=1}^n w_i \frac{D_{(2)i} D_{(2)i}^T}{m_i(1 + (m_i - 1)\phi_i)\pi_2(d_i)(1 - \pi_2(d_i))},$$

with

$$D_{(j)i} = \frac{\partial}{\partial \theta_j} \pi_j(d_i) = (1 - \pi_j(d_i)) \begin{pmatrix} 1 \\ d_i^{\gamma_j} \\ b_j d_i^{\gamma_j} \ln(d_i) \end{pmatrix}, \qquad j = 1, 2.$$

Note that $M(\xi, \theta)$ is a block-diagonal matrix and, as a consequence, the optimality criterion minimizing the variance of the estimate for $ED_\alpha$ can be interpreted as a composite optimality criterion in the sense of Läuter (1974), that is,

$$\begin{aligned}
\mathrm{Var}(\widehat{ED}_\alpha) &\approx \Phi(\xi, \theta) = \bar{D}^T M^{-1}(\xi, \theta) \bar{D} \\
&= \tilde{D}_{(1)}^T M_1^{-1}(\xi, \theta) \tilde{D}_{(1)} + \tilde{D}_{(2)}^T M_2^{-1}(\xi, \theta) \tilde{D}_{(2)}.
\end{aligned} \qquad (4.1)$$

It is intuitively clear that locally optimal designs for estimating the $ED_\alpha$ for overall toxicity are three-point designs if the parameters in $\pi_1(d)$ and $\pi_2(d)$ are similar. In all cases of practical interest these designs have to be calculated numerically. Some examples of optimal designs are presented in Table 4 for constant functions $m_i$ and $\phi_i$. Table 5 presents optimal designs for the case where the correlation is of the form (3.4) and the $m_i$'s are constants. We observe that in most cases, the locally optimal designs are supported at three points, but there are also situations (in particular for large differences between the parameters $\gamma_1$ and $\gamma_2$), where four different dose levels are required for the optimal estimation of the effective dose of the overall toxicity. The results of our investigation of the robustness properties of the locally optimal designs for estimating the effective dose with respect to misspecification of the initial parameters are summarized in Table 6.

We next investigate whether the locally optimal design for estimating the effective dose of prenatal death is efficient for estimating the effective dose of overall toxicity. We also compare the optimal design with the equidistant design with five different dose levels. In Tables 7 and 8, we display efficiencies of the two designs for various combinations of the parameter $\theta$ to study their robustness for estimating the effective dose of overall toxicity when the initial parameters have been misspecified and the design is optimal for estimating the effective dose of prenatal death.

We observe that the performance of the locally optimal design for estimating the effective dose of overall toxicity depends sensitively on changes of the parameters $b_1$ and $b_2$. If $b_1$ is very different from the parameter $b_2$ used in the construction of the locally optimal design for estimating the effective dose of prenatal death, this design becomes inefficient for estimating overall toxicity. In such cases even the uniform design performs better. Otherwise the locally optimal design for estimating the effective dose of prenatal death is at least as good as the uniform design (and in many cases substantially better). The table also shows efficiencies of the design $\xi_{mm}$, which will be constructed in the following section as a robust and efficient alternative to locally optimal designs. The design $\xi_{mm}$ performs uniformly better than the locally optimal design $\xi^*(\theta_0)$ for estimating the effective dose of prenatal death. In many cases, it is substantially more efficient than the uniform design $\xi_u$ and in the cases where the equal allocation rule yields the best efficiencies, the loss of efficiency obtained from $\xi_{mm}$ is rather small.

**Table 4.** Locally optimal designs for estimating the effective dose of overall toxicity conditional on the number of implants. The functions $m$ and $\phi$ are constant, $\alpha = 0.05$ and $\xi_u$ denotes the equidistant design with five different dose levels $0, 1/4, 1/2, 3/4, 1$

| $a_1$ | $b_1$ | $\gamma_1$ | $a_2$ | $b_2$ | $\gamma_2$ | $d_1$ | $d_2$ | $d_3$ | $d_4$ | $w_1$ | $w_2$ | $w_3$ | $w_4$ | $\text{eff}(\xi_u)$ |
|---|---|---|---|---|---|---|---|---|---|---|---|---|---|---|
| 0.06 | 0.7 | 2    | 0.13 | 0.15 | 2    | 0 | 0.495 | 1     |   | 0.330 | 0.546 | 0.124 |       | 0.653 |
| 0.06 | 0.7 | 2    | 0.13 | 0.15 | 3.33 | 0 | 0.493 | 1     |   | 0.291 | 0.574 | 0.134 |       | 0.699 |
| 0.06 | 0.7 | 3.37 | 0.13 | 0.15 | 2    | 0 | 0.573 | 1     |   | 0.372 | 0.536 | 0.092 |       | 0.593 |
| 0.06 | 0.7 | 3.37 | 0.13 | 0.15 | 3.33 | 0 | 0.658 | 1     |   | 0.331 | 0.546 | 0.123 |       | 0.634 |
| 0.06 | 0.5 | 3.37 | 0.13 | 0.3  | 3.33 | 0 | 0.665 | 1     |   | 0.333 | 0.549 | 0.118 |       | 0.607 |
| 0.06 | 0.7 | 3.37 | 0.13 | 0.3  | 3.33 | 0 | 0.653 | 1     |   | 0.321 | 0.551 | 0.128 |       | 0.619 |
| 0.06 | 0.9 | 3.37 | 0.13 | 0.3  | 3.33 | 0 | 0.640 | 1     |   | 0.311 | 0.551 | 0.138 |       | 0.630 |
| 0.06 | 0.7 | 3.37 | 0.05 | 0.3  | 3.33 | 0 | 0.630 | 1     |   | 0.299 | 0.577 | 0.124 |       | 0.593 |
| 0.06 | 0.7 | 3.37 | 0.25 | 0.3  | 3.33 | 0 | 0.673 | 1     |   | 0.338 | 0.532 | 0.130 |       | 0.635 |
| 0.02 | 0.7 | 3.37 | 0.13 | 0.3  | 3.33 | 0 | 0.646 | 1     |   | 0.315 | 0.559 | 0.126 |       | 0.641 |
| 0.09 | 0.7 | 3.37 | 0.13 | 0.3  | 3.33 | 0 | 0.657 | 1     |   | 0.325 | 0.546 | 0.129 |       | 0.613 |
| 0.02 | 1.2 | 2.2  | 0.05 | 0.2  | 3.7  | 0 | 0.402 | 0.636 | 1 | 0.212 | 0.620 | 0.040 | 0.129 | 0.655 |
| 0.02 | 1.2 | 2.2  | 0.05 | 0.2  | 3.3  | 0 | 0.421 | 0.541 | 1 | 0.221 | 0.596 | 0.041 | 0.141 | 0.680 |
| 0.02 | 0.9 | 2.2  | 0.05 | 0.2  | 3.7  | 0 | 0.434 | 0.590 | 1 | 0.228 | 0.585 | 0.065 | 0.121 | 0.674 |
| 0.02 | 1.6 | 2.2  | 0.05 | 0.2  | 3.7  | 0 | 0.368 | 0.713 | 1 | 0.198 | 0.636 | 0.029 | 0.136 | 0.632 |

**Table 5.** Locally optimal designs for estimating the effective dose of overall toxicity conditional on the number of implants. The functions $m$ is constant, while the correlation function $\phi$ is given by (3.4), $a_1 = 0.06$, $b_1 = 0.7$, $\gamma_1 = 3.37$, $a_2 = 0.13$, $b_2 = 0.3$, $\gamma_2 = 3.33$, $\alpha = 0.05$ and $\xi_u$ denotes the equidistant design with five different dose levels $0, 1/4, 1/2, 3/4, 1$

| $u_1$ | $u_2$ | $d_1$ | $d_2$ | $d_3$ | $w_1$ | $w_2$ | $w_3$ | $\text{eff}(\xi_u)$ |
|---|---|---|---|---|---|---|---|---|
| 0 | $-1$ | 0 | 0.600 | 1 | 0.227 | 0.652 | 0.121 | 0.554 |
| 0 | $-2$ | 0 | 0.594 | 1 | 0.192 | 0.690 | 0.118 | 0.538 |
| 0 | $-3$ | 0 | 0.596 | 1 | 0.174 | 0.708 | 0.118 | 0.530 |

## 5. Robust and efficient designs for prenatal death

As pointed out in the previous sections, locally optimal designs are not necessarily robust with respect to a misspecification of the unknown parameters. To obtain designs that are efficient and robust over a certain range of the parameters for the Weibull model, we study a maximin approach proposed by Müller (1995) and Dette (1997), which assumes that there is prior information on the range of plausible values of unknown parameters. To be precise, we concentrate on optimal designs for estimating the effective dose of prenatal death, where the correlation function is given by the one parametric logistic

family

$$\phi(d) = \frac{2}{1 + e^{-ud}} - 1. \tag{5.1}$$

We assume that the experimenter has some knowledge about the location of the parameters, that is,

$$a_2 \in [\underline{a}, \overline{a}], \quad b_2 \in [\underline{b}, \overline{b}], \quad \gamma_2 \in [\underline{\gamma}, \overline{\gamma}], \quad u \in [\underline{u}, \overline{u}].$$

For given $\theta^T = (a_2, b_2, \gamma_2)$ and $u$, define $\xi^*(\theta, u)$ as the locally optimal designs for estimating the effective dose and, for a given design, define

$$\text{eff}_{\text{all}}(\xi, \theta, u) \;=\; \frac{\tilde{D}^T M^{-1}(\xi^*(\theta, u), \theta, u)\tilde{D}}{\tilde{D}^T M^{-1}(\xi, \theta, u)\tilde{D}}. \tag{5.2}$$

A design $\xi_{mm}$ is called standardized maximin optimal for estimating the effective dose if it maximizes the worst efficiency over some set of the parameters, that is,

$$\xi_{mm} = \arg\max_{\xi} \min_{(\theta, u) \in \Omega} \text{eff}_{\text{all}}(\xi, \theta, u). \tag{5.3}$$

Here the set $\Omega$ is defined by $\Omega = [\underline{a}, \overline{a}] \times [\underline{b}, \overline{b}] \times [\underline{\gamma}, \overline{\gamma}] \times [\underline{u}, \overline{u}]$ and is user-selected. Optimal designs with respect to this robust criterion have to be calculated numerically in all cases of practical interest. In Table 9 we display standardized maximin optimal designs with respect to various sets $\Omega$ assuming the quantities $m_i$ and the correlation function (5.1) are constants, that is, $\underline{u} = \overline{u} = 0$. We observe that in all situations the standardized maximin optimal designs are supported at three points and they include the largest and smallest doses. However, the results of Braess and Dette (2007) indicate that there will also exist standardized maximin optimal designs for estimating the effective dose with a larger number of support points. The table also contains the minimal efficiency of the standardized maximin optimal design for estimating the effective dose, that is,

$$\min_{(\theta, u) \in \Omega} \text{eff}_{\text{all}}(\xi, \theta, u),$$

**Table 6.** Efficiency for estimating the effective dose of overall toxicity using three designs: $\xi^*(\theta_0)$, the locally optimal design for $\theta_0^T = (a_1, b_1, \gamma_1, a_2, b_2, \gamma_2)^T = (0.06, 0.7, 3.37, 0.13, 0.27, 3.33)$, $\xi_u$ the equidistant design with five different dose levels (including the largest and smallest dose) and the design $\xi_{mm} = \{0, 0.694, 1; 0.349, 0.515, 0.136\}$, which is standardized maximin optimal for estimating prenatal death with $[0.05, 0.2] \times [0.2, 0.4] \times [2.5, 4.5]$

| $a_1$ | 0.03 | 0.03 | 0.03 | 0.03 | 0.09 | 0.09 | 0.09 | 0.09 |
|---|---|---|---|---|---|---|---|---|
| $b_1$ | 0.4 | 0.4 | 0.9 | 0.9 | 0.4 | 0.4 | 0.9 | 0.9 |
| $\gamma_1$ | 2.7 | 4.2 | 2.7 | 4.2 | 2.7 | 4.2 | 2.7 | 4.2 |
| $a_2$ | 0.05 | 0.05 | 0.05 | 0.05 | 0.2 | 0.2 | 0.2 | 0.2 |
| $b_2$ | 0.2 | 0.2 | 0.4 | 0.4 | 0.2 | 0.2 | 0.4 | 0.4 |
| $\gamma_2$ | 2.5 | 4.5 | 2.5 | 4.5 | 2.5 | 4.5 | 2.5 | 4.5 |
| eff($\xi^*(\theta_0)$) | 0.875 | 0.873 | 0.681 | 0.982 | 0.981 | 0.705 | 0.880 | 0.882 |
| eff($\xi_u$) | 0.588 | 0.566 | 0.594 | 0.584 | 0.606 | 0.574 | 0.619 | 0.614 |
| eff($\xi_{mm}$) | 0.730 | 0.965 | 0.531 | 0.942 | 0.906 | 0.871 | 0.743 | 0.984 |

**Table 7.** Efficiency for estimating the effective dose of overall toxicity. $\xi^*(\theta_0) = \{0, 0.686, 1; 0.396, 0.548, 0.056\}$: locally optimal design for estimating the effective dose of prenatal death ($\theta_0^T = (a_2, b_2, \gamma_2)^T = (0.13, 0.27, 3.33)$, constant correlation), $\xi_u$ equidistant design with five different dose levels $0, 1/4, 1/2, 3/4, 1$ and design $\xi_{mm} = \{0, 0.694, 1; 0.349, 0.515, 0.136\}$, which is standardized maximin optimal for estimating the effective dose of prenatal death with respect to $\Omega = [0.05, 0.2] \times [0.2, 0.4] \times [2.5, 4.5]$

| $a_1$ | 0.02 | 0.02 | 0.02 | 0.02 | 0.1 | 0.1 | 0.1 | 0.1 |
|---|---|---|---|---|---|---|---|---|
| $b_1$ | 0.3 | 0.3 | 1.1 | 1.1 | 0.3 | 0.3 | 1.1 | 1.1 |
| $\gamma_1$ | 2.5 | 4.5 | 2.5 | 4.5 | 2.5 | 4.5 | 2.5 | 4.5 |
| eff($\xi^*(\theta_0)$) | 0.762 | 0.977 | 0.269 | 0.883 | 0.815 | 0.977 | 0.328 | 0.899 |
| eff($\xi_u$) | 0.669 | 0.583 | 0.770 | 0.611 | 0.618 | 0.594 | 0.644 | 0.598 |
| eff($\xi_{mm}$) | 0.901 | 0.986 | 0.437 | 0.984 | 0.935 | 0.980 | 0.520 | 0.995 |
| $a_1$ | 0.03 | 0.03 | 0.03 | 0.03 | 0.09 | 0.09 | 0.09 | 0.09 |
| $b_1$ | 0.4 | 0.4 | 0.9 | 0.9 | 0.4 | 0.4 | 0.9 | 0.9 |
| $\gamma_1$ | 2.7 | 4.2 | 2.7 | 4.2 | 2.7 | 4.2 | 2.7 | 4.2 |
| eff($\xi^*(\theta_0)$) | 0.754 | 0.959 | 0.443 | 0.885 | 0.796 | 0.965 | 0.491 | 0.899 |
| eff($\xi_u$) | 0.652 | 0.589 | 0.708 | 0.609 | 0.622 | 0.592 | 0.646 | 0.599 |
| eff($\xi_{mm}$) | 0.902 | 0.991 | 0.646 | 0.984 | 0.931 | 0.991 | 0.703 | 0.993 |

**Table 8.** Efficiency for estimating the effective dose of overall toxicity. $\xi^*(\theta_0) = \{0, 0.686, 1; 0.396, 0.548, 0.056\}$: locally optimal design for estimating the effective dose of prenatal death ($\theta_0^T = (a_2, b_2, \gamma_2)^T = (0.13, 0.27, 3.33)$, constant correlation), $\xi_u$ equidistant design with five different dose levels $0, 1/4, 1/2, 3/4, 1$ and design $\xi_{mm} = \{0, 0.694, 1; 0.349, 0.515, 0.136\}$ which is standardized maximin optimal for estimating the effective dose of prenatal death with respect to $\Omega = [0.05, 0.2] \times [0.2, 0.4] \times [2.5, 4.5]$

| $a_1$ | 0.03 | 0.03 | 0.03 | 0.03 | 0.09 | 0.09 | 0.09 | 0.09 |
|---|---|---|---|---|---|---|---|---|
| $b_1$ | 0.4 | 0.4 | 0.9 | 0.9 | 0.4 | 0.4 | 0.9 | 0.9 |
| $\gamma_1$ | 2.7 | 4.2 | 2.7 | 4.2 | 2.7 | 4.2 | 2.7 | 4.2 |
| $a_2$ | 0.05 | 0.05 | 0.05 | 0.05 | 0.2 | 0.2 | 0.2 | 0.2 |
| $b_2$ | 0.2 | 0.2 | 0.4 | 0.4 | 0.2 | 0.2 | 0.4 | 0.4 |
| $\gamma_2$ | 2.5 | 4.5 | 2.5 | 4.5 | 2.5 | 4.5 | 2.5 | 4.5 |
| eff($\xi^*(\theta_0)$) | 0.570 | 0.968 | 0.355 | 0.821 | 0.767 | 0.892 | 0.532 | 0.914 |
| eff($\xi_u$) | 0.588 | 0.566 | 0.594 | 0.584 | 0.606 | 0.574 | 0.619 | 0.614 |
| eff($\xi_{mm}$) | 0.730 | 0.965 | 0.531 | 0.942 | 0.906 | 0.871 | 0.743 | 0.984 |

**Table 9.** Standardized maximin optimal designs for estimating the effective dose of prenatal death conditional on the number of implants. The functions $m$ and $\phi$ are constant, $\alpha = 0.05$ and $\xi_u$ denotes the equidistant design with five different dose levels $0, 1/4, 1/2, 3/4, 1$

| $\underline{a}_2$ | $\overline{a}_2$ | $\underline{b}_2$ | $\overline{b}_2$ | $\underline{\gamma}_2$ | $\overline{\gamma}_2$ | $d_1$ | $d_2$ | $d_3$ | $w_1$ | $w_2$ | $w_3$ | min eff | min eff$(\xi_u)$ |
|---|---|---|---|---|---|---|---|---|---|---|---|---|---|
| 0.1 | 0.12 | 0.25 | 0.3 | 3.1 | 3.5 | 0 | 0.681 | 1 | 0.387 | 0.551 | 0.063 | 0.976 | 0.549 |
| 0.1 | 0.15 | 0.25 | 0.3 | 3.1 | 3.5 | 0 | 0.684 | 1 | 0.389 | 0.546 | 0.065 | 0.972 | 0.549 |
| 0.1 | 0.17 | 0.22 | 0.3 | 3 | 3.7 | 0 | 0.696 | 1 | 0.390 | 0.536 | 0.073 | 0.930 | 0.533 |
| 0.08 | 0.18 | 0.21 | 0.33 | 2.6 | 4 | 0 | 0.691 | 1 | 0.371 | 0.524 | 0.105 | 0.809 | 0.514 |
| 0.07 | 0.19 | 0.2 | 0.34 | 2.5 | 4.1 | 0 | 0.690 | 1 | 0.366 | 0.519 | 0.115 | 0.757 | 0.502 |

**Table 10.** Standardized maximin efficient optimal design for prenatal death conditional on the number of implants. The function $m$ is constant, while the correlation function is given by (5.1) and $\alpha = 0.05$. $\xi_u$ denotes the equidistant design with five different dose levels $0, 1/4, 1/2, 3/4, 1$ and various sets are considered in the standardized maximin optimality criterion (5.3), $\Omega_1(\underline{u}, \overline{u}) = [0.07, 0.19] \times [0.19, 0.34] \times [2.5, 4.1] \times [\underline{u}, \overline{u}]$, $\Omega_2(\underline{u}, \overline{u}) = [0.1, 0.12] \times [0.25, 0.3] \times [3.1, 3.5] \times [\underline{u}, \overline{u}]$

|  | $\underline{u}$ | $\overline{u}$ | $d_1$ | $d_2$ | $d_3$ | $d_4$ | $w_1$ | $w_2$ | $w_3$ | $w_4$ | min eff$(\xi^*)$ | min eff$(\xi_u)$ |
|---|---|---|---|---|---|---|---|---|---|---|---|---|
| | 0 | 1 | 0 | 0.469 | 0.721 | 1 | 0.269 | 0.258 | 0.400 | 0.073 | 0.654 | 0.412 |
| $\Omega_1(\underline{u}, \overline{u})$ | 0 | 2 | 0 | 0.460 | 0.722 | 1 | 0.232 | 0.282 | 0.417 | 0.069 | 0.640 | 0.392 |
| | 1 | 2 | 0 | 0.545 | 0.665 | 1 | 0.239 | 0.110 | 0.535 | 0.117 | 0.655 | 0.392 |
| | 0 | 1 | 0 | 0.653 | 1 | | 0.343 | 0.599 | 0.058 | | 0.896 | 0.469 |
| $\Omega_2(\underline{u}, \overline{u})$ | 0 | 2 | 0 | 0.649 | 1 | | 0.327 | 0.615 | 0.057 | | 0.873 | 0.449 |
| | 1 | 2 | 0 | 0.637 | 1 | | 0.254 | 0.700 | 0.046 | | 0.946 | 0.449 |

and the minimal efficiencies of an equidistant design with five different dose levels. Note that the standardized maximin optimal designs yield reasonable efficiencies over the full set $\Omega$ and that the minimal efficiency of the uniform design over this set is substantially smaller. In Table 10 we consider the case where the correlation can be modeled by the function (5.1). We observe that the standardized maximin optimal designs are supported at three or four points and, compared to Table 9, the efficiencies are smaller. This is intuitively clear because we have incorporated more robustness with respect to the assumption of a constant correlation in the construction of efficient designs for estimating the effective dose. Again the equidistant design yields substantially smaller minimal efficiencies compared to the standardized maximin optimal design.

## 6. Efficiency of standard designs and concluding remarks

It is interesting to evaluate the efficiencies of commonly used designs in developmental toxicity studies. One such class is the set of uniform designs. These designs are equally spread out in the dose interval of interest and an equal number of animals is allocated to each dose. As such, they are intuitive and easy to implement. Krewski, Smythe and Fung (2002) provided an overview of experimental designs for 11 developmental toxicity studies conducted under the U.S. National Toxicology Program. In their Table 1, they listed the doses employed in these studies that involved either rabbits, rats or mice. The designs usually have a roughly equal number of animals at each dose and some of their dose levels, after scaling to our protocol interval [0,1], are listed in our Tables 11 and 12. Following Krewski, Smythe and Fung, we call these "standard" designs.

Tables 11 and 12 display the efficiencies of "standard" designs for estimating the prenatal death rates and the overall toxicity rate. We observe that the standard design can perform poorly when model parameters are misspecified. For instance, the efficiencies of the standard design listed in the first row can be less than 30% for estimating the prenatal death rate and the overall toxicity rate. Some standard designs have efficiencies as low as 0.22 for estimating the prenatal death rate. Interestingly, the uniform design with five doses has at least 50% for all cases shown in the tables. The second parts of the two tables show the efficiencies of a uniform design with three support points (note that the local optimal design usually has two or three support points). One observes that this design yields very low efficiencies and cannot be recommended. In general, it is advisable that the researcher assess the efficiencies of a design under different optimality criteria before its implementation.

In practice, there are usually several objectives in the study and these objectives may not be of equal interest to the researcher. For instance, the researcher may be interested in designing a study whose primary aim is to estimate the prenatal death rate, secondary aim is to estimate the malformation rate and tertiary aim is to estimate the overall toxicity rate as accurately as possible. To incorporate the multiple objectives in the study, one may follow the strategy laid out in Cook and Wong (1994) to find an optimal design that provides user-specified efficiency for each objective. Clearly, the optimal design sought should provide higher efficiencies for more important objectives and user-specified efficiencies reasonable enough so that the optimal design exists. For space consideration, we do not provide multiple-objective optimal designs for simultaneously estimating the effective dose for prenatal death rate, malformation rate and overall toxicity rate, but note that the key idea for finding such a design is to first formulate each objective as a convex function of the design information matrix and then combine all the convex objectives into a single convex functional using a convex combination. As described in Cook and Wong (1994), each set of weights used in the convex combination can be judiciously chosen to satisfy the efficiency requirement for each objective. In the case of a two-objective design problem, the weights and the dual-objective optimal design can be determined graphically via efficiency plots. Wong (1999) provided several illustrative applications of such ideas to construct multiple-objective optimal designs in several biomedical problems.

**Table 11.** Efficiency of "standard" designs for estimating $ED_\alpha$ for prenatal death with different values of parameters, $\phi(d) \equiv$ constant

| $d_1$ | $d_2$ | $d_3$ | $d_4$ | $d_5$ | $a_2$<br>$b_2$<br>$\gamma_2$ | 0.13<br>0.27<br>3.3 | 0.05<br>0.27<br>3.3 | 0.13<br>0.15<br>3.3 | 0.13<br>0.27<br>2 | 0.05<br>0.15<br>2 | 0.13<br>0.27<br>1 |
|---|---|---|---|---|---|---|---|---|---|---|---|
| 0 | 0.25  | 0.5  | 1    |   |  | 0.28 | 0.31 | 0.23 | 0.53 | 0.46 | 0.75 |
| 0 | 0.33  | 0.67 | 0.83 | 1 |  | 0.61 | 0.55 | 0.57 | 0.53 | 0.48 | 0.52 |
| 0 | 0.25  | 0.5  | 0.75 | 1 |  | 0.56 | 0.54 | 0.51 | 0.57 | 0.51 | 0.62 |
| 0 | 0.3   | 0.5  | 0.7  | 1 |  | 0.54 | 0.54 | 0.46 | 0.62 | 0.54 | 0.65 |
| 0 | 0.17  | 0.33 | 0.67 | 1 |  | 0.49 | 0.47 | 0.44 | 0.50 | 0.45 | 0.63 |
| 0 | 0.05  | 0.15 | 0.5  | 1 |  | 0.28 | 0.31 | 0.24 | 0.49 | 0.43 | 0.51 |
| 0 | 0.125 | 0.25 | 0.5  | 1 |  | 0.26 | 0.29 | 0.22 | 0.46 | 0.40 | 0.64 |
| 0 | 0.1   | 0.2  | 0.5  | 1 |  | 0.27 | 0.30 | 0.23 | 0.46 | 0.40 | 0.58 |
| 0 | 0.3 | 1 |  |  |  | 0.04 | 0.05 | 0.03 | 0.29 | 0.27 | 0.73 |
| 0 | 0.4 | 1 |  |  |  | 0.14 | 0.18 | 0.11 | 0.52 | 0.45 | 0.71 |
| 0 | 0.5 | 1 |  |  |  | 0.34 | 0.39 | 0.27 | 0.69 | 0.60 | 0.58 |
| 0 | 0.6 | 1 |  |  |  | 0.58 | 0.61 | 0.47 | 0.73 | 0.64 | 0.40 |
| 0 | 0.7 | 1 |  |  |  | 0.74 | 0.69 | 0.64 | 0.59 | 0.53 | 0.23 |

**Table 12.** Efficiency of "standard" designs for estimating $ED_\alpha$ for overall toxicity with different values of parameters in the Weibull model with $\phi(d) \equiv$ constant

| $d_1$ | $d_2$ | $d_3$ | $d_4$ | $d_5$ | $a_1$<br>$b_1$<br>$\gamma_1$<br>$a_2$<br>$b_2$<br>$\gamma_2$ | 0.06<br>0.7<br>3.37<br>0.13<br>0.3<br>3.33 | 0.06<br>0.7<br>3.37<br>0.05<br>0.3<br>3.33 | 0.06<br>0.7<br>3.37<br>0.13<br>0.1<br>3.33 | 0.06<br>0.7<br>3.37<br>0.13<br>0.3<br>1 | 0.06<br>0.7<br>1<br>0.13<br>0.3<br>3.33 | 0.06<br>0.2<br>3.37<br>0.13<br>0.3<br>3.33 | 0.02<br>0.7<br>3.37<br>0.13<br>0.3<br>3.33 |
|---|---|---|---|---|---|---|---|---|---|---|---|---|
| 0 | 0.25  | 0.5  | 1    |   |  | 0.36 | 0.39 | 0.34 | 0.78 | 0.69 | 0.29 | 0.37 |
| 0 | 0.33  | 0.67 | 0.83 | 1 |  | 0.61 | 0.56 | 0.64 | 0.53 | 0.40 | 0.63 | 0.63 |
| 0 | 0.25  | 0.5  | 0.75 | 1 |  | 0.62 | 0.59 | 0.64 | 0.64 | 0.55 | 0.59 | 0.64 |
| 0 | 0.3   | 0.5  | 0.7  | 1 |  | 0.63 | 0.62 | 0.64 | 0.66 | 0.52 | 0.57 | 0.65 |
| 0 | 0.17  | 0.33 | 0.67 | 1 |  | 0.54 | 0.51 | 0.55 | 0.65 | 0.69 | 0.52 | 0.55 |
| 0 | 0.05  | 0.15 | 0.5  | 1 |  | 0.35 | 0.38 | 0.34 | 0.53 | 0.59 | 0.30 | 0.36 |
| 0 | 0.125 | 0.25 | 0.5  | 1 |  | 0.33 | 0.35 | 0.31 | 0.67 | 0.76 | 0.27 | 0.34 |
| 0 | 0.1   | 0.2  | 0.5  | 1 |  | 0.34 | 0.36 | 0.33 | 0.61 | 0.72 | 0.29 | 0.35 |
| 0 | 0.3   | 1    |      |   |  | 0.05 | 0.06 | 0.05 | 0.75 | 0.67 | 0.04 | 0.06 |
| 0 | 0.4   | 1    |      |   |  | 0.19 | 0.23 | 0.18 | 0.72 | 0.51 | 0.15 | 0.21 |
| 0 | 0.5   | 1    |      |   |  | 0.45 | 0.50 | 0.42 | 0.58 | 0.34 | 0.36 | 0.47 |
| 0 | 0.6   | 1    |      |   |  | 0.71 | 0.73 | 0.70 | 0.39 | 0.20 | 0.63 | 0.72 |
| 0 | 0.7   | 1    |      |   |  | 0.78 | 0.72 | 0.79 | 0.22 | 0.11 | 0.78 | 0.76 |

One may be rightly concerned that the optimal designs are dependent on the parametric models. This dependence is inescapable but as we have advocated all along, the user must check robustness properties of optimal designs to all assumptions before the design is implemented. We focused on the simpler situation when we were concerned about misspecification of initial values, but if there is concern about other aspects in the model assumptions, a similar strategy can be applied. For instance, one may question the validity of the Weibull models to describe the malformation and prenatal death rates. If scientific opinion suggests alternative models may be more appropriate, one can then construct optimal designs for different models and compare their efficiencies under the competing models. The hope is that there is a design that remains efficient under all models on which experts may agree.

Here is a short illustration of the situation just discussed: Assume, as before, that both the malformation and prenatal death rates have the same form and can be described using two plausible models:

$$\pi_2^{(2)}(d) = 1 - \frac{a_2}{1+b_2 d^{\gamma_2}}$$

and

$$\pi_2^{(3)}(d) = 1 - \frac{a_2}{1+\mathrm{e}^{-b_2+\gamma_2 d}}.$$

Suppose the sets of initial values are $\theta_2^{(2)} = (0.88, 0.25, 2.8)$, $\theta_2^{(3)} = (0.91, 4.3, 3.5)$, $\theta_1^{(2)} = (0.94, 1.3, 5.1)$ and $\theta_1^{(3)} = (0.98, 3.5, 3.2)$. We recall that $\theta_1^{(1)} = (0.06, 0.7, 3.37)$ and $\theta_2^{(1)} = (0.13, 0.27, 3.33)$. Here the superscripts denote the three different models used to describe the probability rates.

Table 13 lists the locally optimal designs for $\alpha = 0.05$ and their efficiencies under different assumptions on the probability models. The robustness properties of each optimal design under each set of probability models can be compared. For this setup, the efficiency results are quite reassuring because the smallest efficiency in the table is at least 0.76. Of course, different assumptions on the sets of initial values may not yield the same conclusions.

In summary, our proposed design strategy is quite general and possess several advantages over existing methods. Unlike uniform designs, our approach is based firmly on statistical principles and the proposed maximin optimal design provides good protection against misspecification in the initial values of the model parameters. The optimal design allows prior information to be included in its construction and, if required, can also incorporate multiple objectives with possibly unequal interests. Consequently, the proposed optimal design is able to meet the practical needs of the researcher more adequately than current designs.

## Appendix: Proofs

**Proof of Theorem 1.** From (3.1), the information matrix for a design $\xi$ can be represented as

$$M(\xi, \theta) = \sum_{i=1}^{k} w_i f(d_i) f^T(d_i),$$

**Table 13.** Various locally optimal designs (leftpart) and their efficiencies under different probability models for estimating of prenatal death (first three rows), malformation rate (middle three rows) and overall toxicity (last three rows)

| Model | $x_1$ | $x_2$ | $x_3$ | $w_1$ | $w_2$ | $w_3$ | $\xi^{(1)}$ | $\xi^{(2)}$ | $\xi^{(3)}$ |
|---|---|---|---|---|---|---|---|---|---|
| weibull | 0 | 0.686 | 1 | 0.396 | 0.548 | 0.056 | 1.000 | 0.910 | 0.869 |
| model 2 | 0 | 0.624 | 1 | 0.417 | 0.546 | 0.037 | 0.911 | 1.000 | 0.968 |
| model 3 | 0 | 0.630 | 1 | 0.354 | 0.561 | 0.086 | 0.853 | 0.940 | 1.000 |
| weibull | 0 | 0.616 | 1 | 0.297 | 0.602 | 0.101 | 1.000 | 0.954 | 0.832 |
| model 2 | 0 | 0.662 | 1 | 0.284 | 0.592 | 0.123 | 0.918 | 1.000 | 0.503 |
| model 3 | 0 | 0.535 | 1 | 0.290 | 0.610 | 0.100 | 0.882 | 0.768 | 1.000 |
| weibull | 0 | 0.654 | 1 | 0.323 | 0.550 | 0.127 | 1.000 | 0.885 | 0.726 |
| model 2 | 0 | 0.581 | 1 | 0.357 | 0.521 | 0.121 | 0.825 | 1.000 | 0.910 |
| model 3 | 0 | 0.544 | 1 | 0.297 | 0.564 | 0.138 | 0.761 | 0.942 | 1.000 |

where the vector $f$ is defined by

$$f(d) = \frac{D(d)}{\sqrt{m(1+(m-1)\phi(d))\pi_2(d)(1-\pi_2(d))}}$$

$$= \sqrt{\frac{(1-\pi_2(d))}{m(d)(1+(m(d)-1)\phi(d))\pi_2(d)}} \begin{pmatrix} 1 \\ d^{\gamma_2} \\ b_2 d^{\gamma_2} \ln(d) \end{pmatrix}.$$

We now apply Elfving's theorem (see Elfving (1952)), which gives a geometric characterization of the optimal design. More precisely, from this result it follows that a design $\xi = \{d_i, w_i\}_{i=1}^k$ is locally optimal if and only if there exist numbers $\varepsilon_1, \ldots, \varepsilon_k \in \{-1, 1\}$ such that for some $\nu \in \mathbb{R}$ the point

$$\nu P = \nu \left(0, 1/b_2, \frac{1}{\gamma_2}\ln\left(-\frac{\ln(1-\alpha)}{b_2}\right)\right)^T = \sum_{j=1}^k \varepsilon_j w_j f(d_j) \qquad (A.1)$$

is a boundary point of the Elfving set

$$\mathcal{R} = \text{conv}(\{\varepsilon f(d) \mid d \in [0,1], \varepsilon \in \{-1,1\}\}). \qquad (A.2)$$

A typical picture of this set is presented in Figure A.1 for the case of constant functions $\phi$ and $m$. We note that the curve

$$\mathcal{X} = \{f(d), d \in [0,1]\}$$

is contained in subspace $\{x = (x_1, x_2, x_3)^T \in \mathbb{R}^3 \mid x_1 > 0\}$ and the set

$$\{(1, d^{\gamma_2}, b_2 d^{\gamma_2} \ln(d)) \mid d \in [0,1]\}$$

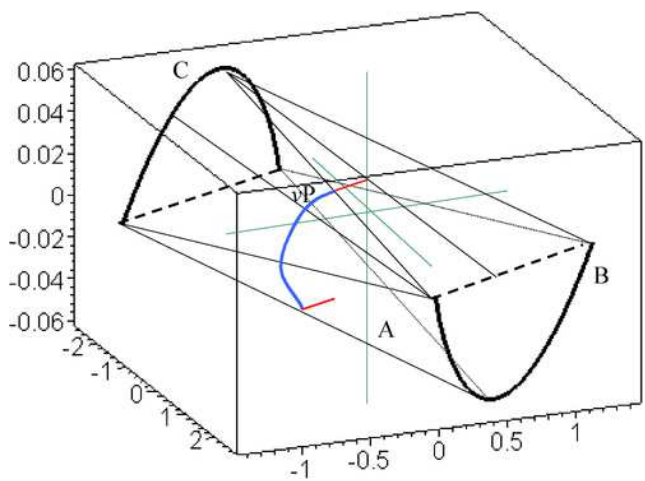

**Figure A.1.** The Elfving set defined in (A.2) for the parameters $a = 0.133$, $b = 0.272$, $\gamma = 3.33$. The points $f(d_1)$, $-f(d_2)$ and $f(d_3)$ are denoted by $A$, $C$ and $B$, respectively, while the point $\nu P$ is defined in (A.1).

defines a U-shaped curve. From the monotonicity assumption for the function (3.3), it follows that the curve $\mathcal{X}$ is also U-shaped (see also Figure A.1). We denote the end-points of this curve by $A$ and $B$ and recall that the first coordinate of the vector $P$ is equal to 0 and that $\nu$ is the scaling constant such that $\nu P$ touches the boundary of the Elfving set $\mathcal{R}$. Note that in the case $\alpha \to 0$ we have that

$$P \approx c \begin{pmatrix} 0 \\ 0 \\ 1 \end{pmatrix}$$

for some constant $c$ and, consequently, the vector $\nu P$ touches the boundary at the plane $\mathcal{E}$ spanned by the points $A$, $B$ and $C$, where $A$ and $B$ correspond to the doses 0 and 1, respectively, and $-C$ corresponds to a third dose, say $d^* \in (0, 1)$. Consequently, the locally optimal design is a three-point design with support points 0, 1 and $d^*$, if $\alpha$ is sufficiently small. In the case where $\alpha \to 1$, the situation is exactly the same and the locally optimal design is also supported at three points, including the boundary points. From the geometry of the Elfving set $\mathcal{R}$ we see that there is also direction $P$, where the intersection with the Elfving set can be represented by two points of the curves $\mathcal{X}$ and $-\mathcal{X}$. In particular, this situation occurs if $\alpha \approx 1 - \mathrm{e}^{-b}$. In this case we have

$$P \approx c \begin{pmatrix} 0 \\ 1 \\ 0 \end{pmatrix}$$

for some constant $c$ and the locally optimal design is supported by a two-point design. Moreover, if $\alpha$ moves from 0 to 1, it follows from the geometry of the Elfving set that the situation is changing continuously, which proves the assertion of the theorem. $\square$

**Proof of Theorem 2.** Multiplying the last coordinate of the vector equation (A.1) by $\gamma_2$ yields

$$\nu P = \nu \begin{pmatrix} 0 \\ 1/b_2 \\ \ln\left(-\dfrac{\ln(1-\alpha)}{b_2}\right) \end{pmatrix}$$

$$= \sum_i \varepsilon_i w_i \sqrt{\dfrac{(1-\pi_2(d_i))}{m(1+(m-1)\phi)\pi_2(d_i)}} \begin{pmatrix} 1 \\ d_i^{\gamma_2} \\ b_2 d_i^{\gamma_2} \ln(d_i^{\gamma_2}) \end{pmatrix} \qquad (A.3)$$

for the boundary point $\nu P \in \mathcal{R}$. If $\{d_i^*(1); w_i^*\}$ denotes an optimal design for the parameter $\theta = (a_2, b_2, 1)$, it follows that equation (A.3) holds for this design with $\gamma_2 = 1$. Now it is easy to see that (A.3) is also true for the design $\{(d_i^*(1))^{1/\gamma_2}; w_i^*\}$ for the parameter $\theta = (a_2, b_2, \gamma_2)$, where $\gamma_2 > 0$ is arbitrary. This proves the first part of the theorem. The second part follows by similar arguments, which are omitted by the sake of brevity. $\square$

**Proof of Theorem 3.** Part (a) of the Theorem follows directly from the geometry of the Elfving set. If the locally optimal design is supported at three points, the corresponding point $\nu P$ touches the Elfving set in the plane spanned by the points $A$, $B$ and $C$, which does not depend on the value of $\alpha$. For a proof of part (b) we note that, according to Elfvings theorem, a locally optimal design of the form $\{0, d_2, w_1, w_2\}$ must satisfy the equation

$$\nu \begin{pmatrix} 0 \\ 1/b_2 \\ \ln\left(-\dfrac{\ln(1-\alpha)}{b_2}\right) \end{pmatrix} = \varepsilon w_1 g(0) \begin{pmatrix} 1 \\ 0 \\ 0 \end{pmatrix} - \varepsilon w_2 g(d_2) \begin{pmatrix} 1 \\ d_2^{\gamma_2} \\ b_2 d_2^{\gamma_2} \ln(d_2^{\gamma_2}) \end{pmatrix},$$

where the function $g$ is defined by

$$g(d) = \sqrt{\dfrac{(1-\pi_2(d))}{m(d)(1+(m(d)-1)\phi(d))\pi_2(d)}} \, .$$

It is easy to see that this equation yields

$$\nu \begin{pmatrix} 1/b_2 \\ \ln\left(-\dfrac{\ln(1-\alpha)}{b_2}\right) \end{pmatrix} = -\varepsilon w_2 \sqrt{\dfrac{(1-\pi_2(d_2))}{m(d_2)(1+(m(d_2)-1)\phi(d_2))\pi_2(d_2)}} \begin{pmatrix} d_2^{\gamma_2} \\ b_2 d_2^{\gamma_2} \ln(d_2^{\gamma_2}) \end{pmatrix},$$

which simplifies to the equation

$$\nu\begin{pmatrix}1\\\ln(ED_\alpha^{\gamma_2})\end{pmatrix} = -\varepsilon w_2\sqrt{\frac{(1-\pi_2(d_2))}{m(d_2)(1+(m(d_2)-1)\phi(d_2))\pi_2(d_2)}}b_2 d_2^{\gamma_2}\begin{pmatrix}1\\\ln(d_2^{\gamma_2})\end{pmatrix}.$$

It follows that $d_2 = ED_\alpha$. Since $w_1 = 1 - w_2$ from the equality for the first coordinate, we have that $w_2 = g(0)/(g(0) + g(ED_\alpha))$. □

## Acknowledgements

The support of the Deutsche Forschungsgemeinschaft (SFB 475, "Komplexitätsreduktion in multivariaten Datenstrukturen") is gratefully acknowledged. The work of H. Dette and W.K. Wong was supported in part by NIH Grant award IR01GM072876. Additionally, Wong was supported by NIH Grant P30 CA16042-33.